\documentstyle[preprint,aps]{revtex}
\Roman{section}

\begin{document}
\draft
\preprint{BIHEP-TH-96-18}

\title{STUDY ON $\nu_{\mu}$, $\nu_{\tau}$ DECAYS AND CP-NON
INVARIANCE AT EARLY EPOCH OF THE UNIVERSE\footnote{This 
work is partly supported by the National Science
Foundation of China (NSFC) }}
\maketitle

\centerline{Wu-Sheng Dai$^{1}$, Song Gao$^{2}$
and Xue-Qian Li$^{1}$ }
\vspace{0.4cm}
\begin{center}
{\small
1. Department of Physics, Nankai University,

Tianjin, 300071, China.
\vspace{8pt}

2. Institute of High Energy Physics, Academia Sinica,

P. O. Box 918 (4),  Beijing 100039, China.}

\end{center}
\vspace{1cm}

\begin{abstract}
We study the $\nu_{\mu}$ and $\nu_{\tau}$ decays in the early epoch of
the universe. If $m_{\nu_{\tau}}>2m_e$, there would be a CP asymmetry
between $\nu_{\tau}\rightarrow e^++e^-+\nu_e$ and $\bar{\nu}_{\tau}
\rightarrow e^++e^-+\bar{\nu}_e$. The resultant CP non-invariance is
a function of temperature and density and can reach $10^{-7}$ for a
reasonable
temperature range, but it is noticed that if $m_{\nu_{\tau}}>2m_{\mu}
\sim 200$ MeV, the CP non-invariance can be much larger.
\end{abstract}


\newpage
\baselineskip 22pt

\noindent {\bf I. Introduction.}

As a possible candidate of the hot dark matter, the characteristics of
neutrinos need to obey many constraints from both cosmology and
experiments on the Earth. The recent research indicates that the upper
limit of the dark matter neutrinos must be less than 10 eV \cite{Morrison}.
The earth experiments give an upper bound on $\nu_e$ mass as $m_{\nu_e}<7.2$
eV \cite{Wein}, the measurements by a PSI group \cite{PSI} has given an
upper limit of $\nu_{\mu}$ as $m_{\nu_{\mu}}<160$ keV. For $\nu_{\tau}$,
there are several groups which have obtained various upper bounds as
\cite{Gentile} in Table I.

\vspace{0.6cm}

\begin{center}

\begin{tabular}{|l|c|c|c|c|}
\hline
Exp.char. & CLEO & ARGUS & OPAL & ALEPH \\
\hline
Produced $\tau^+\tau^-$ & 1.77M & 325K & 36K & 76K \\
signal & $3\pi^{\pm}\pi^0/5\pi^{\pm}$ & $5\pi^{\pm}$ & $5\pi^{\pm}(\pi^0)$
& $5\pi^{\pm}/5\pi^{\pm}\pi^0$ \\
2$^{\rm nd}$ $\tau$ & 53/60 & 20 & 5 &23/2 \\
Method & 1--D & 1--D & 2--D & 2--D \\
\hline
$m_{\nu_e}$ ($<$95\% CL) & 32.6 & 31 & 74 & 23.8        \\
\hline
\end{tabular}

\vspace{0.3cm}

Table I. The mass is in unit of MeV.

\end{center}

\vspace{0.8cm}

The stringent constraints demand that either the neutrino as the hot dark
matter component is lighter than 10 eV, or can decay fast enough, so that
its relic contents are not substantial in the present universe. Of course,
there is possibility for existence of heavy unstable neutrinos to which
the cosmology sets lower bounds \cite{Lee}. But a crucial problem is to
investigate the decays of the lighter neutrinos $\nu_{\mu}$ and
$\nu_{\tau}$.

The cosmology demands that corresponding to a few MeV, the minimum lifetime
is $\tau\leq 2.5\times 10^8$ sec. for Dirac neutrinos and $\tau\leq 4.3\times
10^8$ sec. for Majorana neutrinos \cite{Masso}.

As long as $\nu_{\tau}$ is heavier than $2m_e\sim 1.0$ MeV, it can decay via
$\nu_{\tau}\rightarrow\nu_e+e^++e^-$ and the decay rate at tree level can
be written as
\begin{equation}
\Gamma={G_F^2m_{\nu_{\tau}}^5\over 192\pi^3}|U_{\nu_{\tau}e}^*U_{\nu_ee}|^2,
\end{equation}
where the electroweak (EW) corrections and the mass effects of $m_e$ are 
omitted,
$U_{\nu_{\tau}e}$, $U_{\nu_ee}$ are the Cabibbo-Kobayashi-Maskawa matrix
elements. Provided $m_{\nu_{\tau}}\sim 31$ MeV and $U_{\nu_{\tau}e}\sim 0.01$,
$U_{\nu_ee}\sim 1$, the lifetime $\tau$ can be about 11 sec. Therefore if
$m_{\nu_{\tau}}$ is sufficiently heavy, it can satisfy the cosmology
constraint. However, if $m_{\nu_{\tau}}$ were 30 times lighter (say,
1$\sim$2 MeV), its lifetime would be close to the cosmology bound.

A more serious problem exists for $\nu_{\mu}$, if its mass is about 160 KeV.
Due to the phase space restriction, $\nu_{\mu}$ cannot decay via any
reactions at tree level and the most probable decay mode is $\nu_{\mu}
\rightarrow\nu_{e}\gamma$ which can only occur via the penguin diagram.
Because of the loop suppression and the GIM mechanism, the decay rate would be
very small, if the temperature effects are not taken into account. In fact,
the lifetime can be as long as $10^{22}$ sec. i.e. $3\times 10^{14}$ yrs.

Nieves et al. systematically studied the properties of neutrinos in medium
with finite temperature and density \cite{Nieves}, especially, they
investigated $\nu_{\tau}\rightarrow \nu_e+\gamma$ and found a tremendous
change, because then the GIM suppression is dismissed, so that the decay rate can
be nine orders larger than that at zero temperature \cite{Olivo}.
Following the trend, we will employ
their results to analyze the $\nu_{\mu}$ lifetime, provided $\nu_{\mu}\sim
160$ KeV. We further study the temperature and density effects on the
decay $\nu_{\tau}\rightarrow \nu_e+e^++e^-$. Obviously, the temperature and
density effects occur at the propagators which only exist in the penguin
loop. At zero temperature, the loop correction is small with a factor
$\alpha/\pi\sim 2\times 10^{-3}$ with respect to the tree level. For higher
temperature and density which exist at the early epoch of the universe,
there are two more energy scales besides the mass of $\nu_{\tau}$,
e.g. the temperature $T$ and the electron chemical potential $\mu$ which
corresponds to the density.
Therefore once $T,\; \mu>m_{\nu_{\tau}}$, there is
a chance to elevate the amplitude by orders and this enhancement may
compensate the loop suppression. So we investigate the temperature
and density effects via loops, concretely the penguin diagrams.

The penguin diagram at zero-temperature has been carefully studied by
many authors \cite{Shifman}. Nieves et al. studied the penguin contributions
to $\nu_{\tau}\rightarrow\nu_{e}+\gamma$, where $\gamma$ is a real photon
satisfying on-shell condition $q^2=0$. On contrast, in the process $\nu_{\tau}
\rightarrow\nu_{e}+e^++e^-$, the propagating photon is a virtual
intermediate boson and $q^2\neq 0$, therefore the results would be expected
to deviate from that in $\nu_{\tau}\rightarrow\nu_e+\gamma$. Moreover,
it is well known that as $q^2\neq 0$, the penguin loop may possess an
imaginary part which induces a CP asymmetry for $\nu_{\tau}\rightarrow
\nu_e+e^++e^-$ and $\bar{\nu}_{\tau}\rightarrow \bar{\nu}_e+e^++e^-$
\cite{Du}. The temperature and density effects could modify the CP
non-invariant phenomenon. In fact, any CP non-invariant source in the
early universe is interesting and worth careful studies.

Our paper is organized as following. After the introduction, we give the
necessary formulations, in Sec.III, we present our numerical results,
while the last section is devoted to a brief discussion and conclusion.\\

\noindent II Formulations

(i) For $\nu_{\mu}\rightarrow\nu_e+\gamma$.

Because the final phase space forbids the reaction $\nu_{\mu}\rightarrow
\nu_e+e^++e^-$ as long as $m_{\nu_{\mu}}<1$ MeV, so the most reasonable
decay mode for $\nu_{\mu}$ is the radiative decay $\nu_{\mu}\rightarrow
\nu_e+\gamma$.

Nieves et al. \cite{Nieves}\cite{Olivo} assumed that only electrons
(positrons) in the atmosphere, for the case of a nonrelativistic (NR)
electron background,
\begin{equation}
\label{NR}
{\Gamma^{(NR)}\over\Gamma}=1.39\times^{19}r
F({\cal{V}})\left [{n_e\over 10^{24} cm^{-3}} \right ]^2
\left ({1eV\over m}\right )^4,
\end{equation}

whereas for an extreme relativistic (ER) background,
\begin{equation}
\label{ER}
{\Gamma^{(ER)}\over\Gamma}=1.5\times 10^{9}rF({\cal{V}})
        \left({T\over m}\right )^4
\end{equation}
where $r=|U_{e\nu_e}^*U_{e\nu_{\mu}}|/|U_{\tau\nu_e}^*U_{\tau\nu_{\mu}}|$
which is a large number about $10^{4}\sim 10^{5}$ unlike that in
$\nu_{\tau}\rightarrow \nu_{e}\gamma$ it is close to unity, $\Gamma$ is
the width at zero temperature.
$$F({\cal{V}})=(1-{\cal{V}}^2)^{1/2}\left [{2\over {\cal{V}}}\ln
\left ({1+{\cal{V}}
\over 1-{\cal{V}}}\right )-3\right ]$$
where ${\cal{V}}$ is the three-velocity of the decaying neutrino and
$F({\cal{V}})$ can
take values between 1 and  1.55. The superscripts in eqs.(\ref{NR}, \ref{ER})
denote the non-relativistic and extra-relativistic cases respectively and
$n_e$ is the electron density in the surrounding, $m$ is the mass of $\nu_{\mu}$.

The zero temperature $\Gamma$ is expressed as
\begin{equation}
\Gamma\approx {1\over 2}G_F^2({3\over 32\pi^2})^2m^5\left|\sum_{l=e,\mu,\tau}
{m_l^2\over M_W^2}U_{l\nu_{\mu}}^*U_{l\nu_e}\right|^2.
\end{equation}
Obviously due to the GIM mechanism, only $\tau$ contributes substantially
at zero temperature.

(ii) For $\nu_{\tau}\rightarrow\nu_e+e^++e^-$.

The dispersive part of the amplitude comes from both the tree level and the
the penguin diagrams, while the absorptive part only comes from the penguin.
It is noted that besides the absorptive phase determined by a ratio of
the absorptive part and the dispersive part, there are the
CKM phases and they would be dealt with separately.

The propagator of fermions at finite temperature and density can be
written as [11]
\begin{equation}
S_F(p)=(\rlap/p+m)\left[{1\over p^2-m^2+i\epsilon}+2\pi\delta(p^2-m^2)
\eta(p\cdot v)\right],
\end{equation}
where
\begin{equation}
\eta(x)={\theta(x)\over e^{\beta(x-\mu)}+1}+{\theta(-x)\over e^{-\beta(x-\mu)}
+1},
\end{equation}
and $\beta=1/kT$, $\mu$ is the chemical potential and $v_{\alpha}$ is the
four-velocity of the medium, in the Lab frame $v_{\alpha}=(1,\vec 0)$.

The Feynman diagrams are shown in Fig.1 where (a) corresponds to the tree
diagram, obviously at this energy scale, (c) gives rise of negligible
contributions only. In the real photon emission case where $q^2=0$,
$\delta(p^2-m_e^2)$ and $\delta((p-q)^2-m_e^2)$ cannot be non-zero
simultaneously as long as $m_e^2\neq 0$, therefore, in the reaction
$\nu_{\tau}\rightarrow\nu_e+\gamma$ the penguin diagram does not possess an
absorptive part. However, as  $q^2\neq 0$, product of the
two $\delta-$functions can be non-zero, therefore an absorptive part of the
penguin diagram emerges. For the energy scale of $m_{\nu_{\tau}}\sim 31$ MeV,
only electron-loop contributes an absorptive part, but not $\mu$ and $\tau$,
because $m_{\nu_{\tau}}<2m_{\mu}$. It is a well-known fact in the
zero-temperature field theory, but for the finite temperature and density
situation, their effects on the absorptive part  appear via
$\eta(p\cdot v)\eta((p-q)\cdot v)$.

The calculations for the absorptive part concern an integration
\begin{equation}
\int d^4p\delta(p^2-m_e^2)\delta((p-q)^2-m_e^2)f(p).
\end{equation}
which was investigated by Kobes and Semenoff \cite{Kobes}.

Thus, the total temperature and density dependent amplitude would be
\begin{eqnarray}
M &=& -i\left({g\over 2\sqrt 2}\right)^2\bar u_{\nu_e}\gamma_{\alpha}(1-\gamma_5)
u_{\nu_{\tau}}\bar e\gamma^{\alpha}(1-\gamma_5)e{1\over k^2-M_W^2}U^*_{e\nu_e}
U_{e\nu_{\tau}} \nonumber\\
&+& (-i)\bar u_{\nu_e}\Gamma_{\alpha}^{(eff)}
{1\over q^2}\bar e\gamma^{\alpha}e,
\end{eqnarray}
where $\Gamma^{(eff)}_{\alpha}$ stands for the loop contributions of $e$,
$\mu$ and $\tau$ as
\begin{equation}
\Gamma^{(eff)}_{\alpha}=\sum_{l=e,\mu,\tau}\left(\Gamma_{l\alpha}^{D(eff)}
+i\Gamma_{l\alpha}^{A(eff)}\right )U^*_{l\nu_e}U_{l\nu_{\tau}},
\end{equation}
and the superscripts D and A represent the dispersive and absorptive parts
of the penguin diagram respectively.

In the derivations, we follow Nieves et al. [7] to use their conventions and
decomposition as
\begin{equation}
{\cal{T}}_{\alpha\beta}=M_TR_{\alpha\beta}+M_LQ_{\alpha\beta}+M_PP_{\alpha\beta},
\end{equation}
where
\begin{eqnarray}
R_{\alpha\beta} &\equiv&  \tilde g_{\alpha\beta}-Q_{\alpha\beta},\\
Q_{\alpha\beta} &\equiv&  {\tilde v_{\alpha}\tilde v_{\beta}\over\tilde v^2} \\
P_{\alpha\beta} &\equiv&  {i\over Q^2}\epsilon_{\alpha\beta\lambda\mu\nu}q^{\mu}
v^{\nu},
\end{eqnarray}
and $\tilde g_{\alpha\beta}\equiv g_{\alpha\beta}-{q_{\alpha}q_{\beta}\over
q^2}$, $\tilde v_{\alpha}\equiv \tilde g_{\alpha\beta}v^{\beta}$.

Then the following calculations are straightforward and standard, even though
tedious, we ignore the details in the work to save space.

Thus the amplitude squared can be easily written down in a standard
framework. The decay width is
\begin{eqnarray}
\Gamma^{\rm total} &=& \int {d^3p_{\nu_{\tau}}\over (2\pi)^3}f_{\nu_{\tau}}
(p_{\nu_{\tau}}){1\over 4E_{\nu_{\tau}}}\int {d^3p_{e^+}\over (2\pi)^3}
{1\over 2E_{e^+}}\cdot {d^3p_{e^-}\over (2\pi)^3}{1\over 2E_{e^-}}\cdot
{d^3p_{\nu_e}\over (2\pi)^3}{1\over 2E_{\nu_e}} \nonumber \\
&&\cdot |M|^2\delta^4(p_{\nu_{\tau}}-p_{e^+}-p_{e^-}-p_{\nu_e})
(1-f_{\nu_e})(1-f_{e^+})(1-f_{e^-}),
\end{eqnarray}
where $f_i$'s are the distribution functions of fermions, $(1-f_i)$ denotes
the Pauli blocking of the produced $e^+$, $e^-$ and $\nu_e$.
It is suggested that in the early universe, one may
use ${d^3p\over 2E}\sqrt{-g}$ where $g$ is the determinant of the metric matrix,
instead of ${d^3p\over 2E}$, but here we assume that at this concerned
time period of the universe evolution, the universe is flat enough
and $\sqrt{-g}$ is close
to unity.

(iii) The CP asymmetry.

For $\nu_{\tau}\rightarrow e^++e^-+\nu_e$, the amplitude can be recast into
another form which is more appropriate for the CP asymmetry discussions as
\begin{equation}
M=A_{(0)}'e^{i\delta_e}+A_{(e)}'e^{i\delta_e}e^{i\theta_e}+A_{(\mu)}e^{i
\delta_{\mu}}+A'_{(\tau)}e^{i\delta_{\tau}},
\end{equation}
where
\begin{eqnarray}
A_{(0)}' &=& |A_{(0)}U_{e\nu_e}^*U_{e\nu_{\tau}}|, \\
A_{(e)}' &=& \left|\sqrt{A_{(e)}^{(D)2}+A_{(e)}^{(A)2}}
U_{e\nu_e}^*U_{e\nu_{\tau}}\right |, \\
A_{(\mu)}' &=& |A_{(\mu)}U_{\mu\nu_e}^*U_{\mu\nu_{\tau}}|, \\
A_{(\tau)}' &=& |A_{(\tau)}U_{\tau\nu_e}^*U_{\tau\nu_{\tau}}|,
\end{eqnarray}
and $A_{(0)}$ is the tree level quantity, while $A_{(l)}$'s correspond to
the loop as $l=e,\mu,\tau$ respectively.

The absorptive phase $\theta_e$ which is equivalent to the strong phase in
$K\rightarrow 2\pi$
process,  comes from the absorptive part of the penguin loop as
\begin{equation}
\tan^{-1}\theta_e={A_e^{(A)}\over A_e^{(D)}}, \;\;\;\;\;\; \theta_{\mu}
=\theta_{\tau}=0.
\end{equation}
Since $m_{\nu_{\tau}}<2m_{\mu}$, only the electron loop contributes an
absorptive part according to the Cutkosky cutting rule \cite{Politzer}
and the weak phases $\delta_e,\;\delta_{\mu},\;\delta_{\tau}$ originate from
the CKM phase. Under a CP transformation, the weak phases change signs,
while $\theta_e$ does not. The interferences give rise to a CP asymmetry.

The CP asymmetry is proportional to
\begin{eqnarray}
\label{asym}
a &\equiv & \frac{\int \prod_k (dLIPS_k)(|M|^2-|\overline M|^2)}{\int\prod_k
(dLIPS_k)(|M|^2+|\overline M|^2)} \nonumber \\
&=& \frac{\int \prod_k(dLIPS_k)(\sum_{i,j}2|A'_i||A'_j|\sin (\delta_i-
\delta_j)\sin(\theta_i-\theta_j)}{\int \prod_kd(LIPS_k)[\sum_{i,j}
[|A'_i|^2+2|A'_i||A'_j|\cos(\delta_i-\delta_j)\cos(\theta_i-\theta_j)]},
\end{eqnarray}
where dLIPS$_k$ denotes the Lorentz Invariant Phase Space of final particle
$k$.

The numerator comes from interferences among all the terms (tree level
and $e,\;\mu,\;\tau$ loops), obviously, it is zero unless both the weak
and strong phases of the two terms are different. Therefore in our case,
the numerator is proportional to
\begin{equation}
2|A'_{(e)}|[|A'_{(\mu)}|\sin(\delta_e-\delta_{\mu})+|A'_{(\tau)}|\sin(\delta_e
-\delta_{\tau})]\sin\theta_e.
\end{equation}

It is observed that there is no contribution to the CP asymmetry
from the interference between the tree amplitude and $A'_i$ because they
either have the same weak phase (with $A'_{(e)}$) or the same absorptive
phase (with $A'_{(\mu)}$ and $A'_{(\tau)}$). As some authors suggested that
in the surrounding there were no $\mu$ and $\tau$ leptons, then $A'_{(\mu)}$
and $A'_{(\tau)}$ are the zero temperature amplitudes which are somewhat
smaller than that with higher temperature and density.

We can argue that since $m_{\tau}\sim 1.7$ GeV $\gg m_{\mu}$, one can expect
it to decay very fast, so that there is no $\tau-$lepton in the atmosphere,
but probably a certain amount of muons.
In our later calculations we assume that both $\mu$ and $\tau$ exist when
the reaction takes place. \\

\noindent III. The numerical results.

In the calculations, we study the high temperature and density effects for
the early universe. The studies on the evolution of the universe
suggested that the reasonable temperature and density
limits may be $4\times 10^{12}$ K and $\mu=600$ MeV, so we take them
as the extreme bounds \cite{Turner}. In the calculations we adopt the CKM
matrix elements as given in the data book \cite{Data} and ref.\cite{Peres}.

(i) For the $\nu_{\mu}$ decay, the zero temperature calculations predict
a rather long lifetime as $\tau_{\nu_{\mu}}\sim 3.5\times 10^{12}$ yrs
which is longer than the lifetime of our universe.

The calculations show that at $T=10^{12}$ K, the lifetime of
$\nu_{\mu}$ decreases to $2\times 10^{3}$ sec. But since the temperature of the universe
cannot hold so high for more than a few seconds,
thus one can expect that an average lifetime
of $\nu_{\mu}$ would be $10^{8}$ sec, which roughly coincides with the
cosmology requirements. We will discuss this consequence in the last section.

(ii) For $\nu_{\tau}$ decays.

There are two important decay modes $\nu_{\tau}\rightarrow e^++e^-+\nu_e$
and $\nu_{\tau}\rightarrow \nu_e+\gamma$. At zero temperature, because
the former reaction can be realized via a tree diagram, so its contribution
dominates over the radiative decay mode which can only occur via loops.

The finite temperature and density effects change the characteristics of the
decay modes. The radiative decay was studied by Nieves et al. in detail,
while we investigate the temperature and density effects on
$\nu_{\tau}\rightarrow e^+e^-\nu_e$ in this work.
Because at high temperature and density,
the new energy scale $T$ and the chemical potential $\mu$ would replace
or partially replace the mass term, so can result in a larger decay rate.
Our numerical results indicate that as  $T$ and $\mu$ get large enough,
the loop contribution can be one times larger than the tree contributions.
(see Fig.2).

In fact we have also found the dependence of the decay rate on the density
as $\Gamma$ increases fast with the density increment. It is the main aim
of this work.
The dependence on the chemical potentials are shown in Fig.2, the three
curves correspond to various temperatures.
Its significance would be discussed
in next section.

It is noted that the penguin contributions increase fast with increments of
temperature either, for $\mu=0$, at about $T=6\times 10^{11}$ K,
the tree amplitude and the penguin have the same magnitude, then the penguin
contribution takes over and dominates.

(iii) The CP asymmetry.

One could expect an absorptive part emerging at zero temperature as
an absorptive angle is resulted \cite{Wu}
\begin{equation}
\tan\theta_e=\frac{{G_F\over\sqrt 2}{2\alpha\over 3}}{{G_F\over\sqrt 2}
({-2\alpha\over 3\pi})\ln{m_i^2\over M_W^2}}
\end{equation}
which is about -0.4, but as the temperature is not zero, this ratio
becomes very small that $\theta_e\sim 10^{-6}$. Concrete values are shown in
Table II.

\vspace{0.3cm}

\begin{center}

\begin{tabular}{|c|c|c|c|c|c|c|c|c|}
\hline
$\mu$ & 0 & 50 & 100 & 200 & 300 & 400 & 500 & 600 \\
\hline
$\tan\theta_e (\times 10^{-6}$ & 4.1 & 4.05 & 3.9 & 3.43 & 2.79 & 2.12 & 1.53
& 1.03 \\
\hline
$a(\times 10^{-7}$ & 4.1 & 4.1 & 4.08 & 3.98 & 3.2 & 2.6 & 2.1 & 1.5 \\
\hline
\end{tabular}

\vspace{0.2cm}

Table II. in the table,
$T=4\times 10^{12}$ K, $\mu$ is in MeV.

\end{center}

\vspace{0.6cm}

In fact, in the interaction
\begin{equation}
\label{finite}
  \int d^4p\delta(p^2)\delta((p-q)^2){1\over e^{\beta(E-\mu)}+1}
{1\over e^{\beta(E+\mu)}+1}(2\pi)^2 F(p),
\end{equation}
where we ignore the small electron mass and $F(p)$ is a function of $p$,
the distributions are generally smaller than unity, on contrast, at the zero
temperature, it is
\begin{equation}
\label{zero}
\int d^4 \delta(p^2)\delta((p-q)^2) (\pi)^2 F(p),
\end{equation}
so it can be expected that the finite temperature and density effects
cannot make the integration values to be much larger than the zero-temperature
case. Especially, as the chemical potential $\mu$ gets larger, the second
distribution in (\ref{finite}) would decrease very steeply, thus the
integration turns smaller accordingly and so does $\tan\theta_e$.
Therefore, we should think that the absorptive
part remains at the same order as the zero temperature value. Our careful
numerical computation confirms it. Thus a rapid increase of the dispersive
part of the penguin loop with temperature and density causes $\tan\theta_e$
to decrease fast and the effects disfavor the CP asymmetry. We will discuss
it in next section.  \\

\noindent IV. Discussion and conclusion.

We have evaluated the temperature and density effects on the $\nu_{\mu}$
and $\nu_{\tau}$ lifetimes and estimated a possible CP asymmetry $a$
for various $T$ and $\mu$.

(i) For the $\nu_{\mu}$ lifetime, it can only decay via the radiative
mode, we estimate its lifetime as about $10^{3}$ sec. as $m_{\nu_{\mu}}=160$
KeV for very high temperature.
However, obviously, such high temperature cannot hold for so long and would
decrease very fast, and so does the $\nu_{\mu}$ decay rate since it
is proportional to $T^4$. Considering this, $\nu_{\mu}$ lifetime is close
to $10^{8}$ sec. which is the bound set by cosmology. It is possible that
as $\nu_{\mu}\leq 160$ KeV, its lifetime bound would be even larger, then
one may conclude that to be consistent
with the cosmology, either the $m_{\nu_{\mu}}$ is much lower than the upper
bound to be close to 10 eV, or there is some new physics which speeds
the decay, for example axion exists etc.

(ii) The temperature and density effects on the $\nu_{\tau}$ lifetime.
The process $\nu_{\tau}\rightarrow \nu_{e}+\gamma$ was carefully studied by
Nieves et al. \cite{Nieves}\cite{Olivo}. We study the effects on $\nu_{\tau}
\rightarrow e^+e^-\nu_{e}$ for which there is a tree diagram. For
$\nu_{\tau}\rightarrow e^+e^-\nu_e$, the temperature and density effects
occur via the penguin loop which is $10^{-3}-10^{-4}$ orders smaller than
the tree amplitude at zero temperature. The numerical results indicate that
the dispersive loop contribution is enhanced to the same order as the tree
amplitude. Moreover,
at $T=10^{12}$ K, the two decay modes are of the
same orders

We have found that as temperature increases, the rates of both modes increase,
below a certain temperature (about
$T=10^{12}$ K) the rate of $\nu_{\tau}\rightarrow e^+e^-\nu_{e}$
is larger than that of $\nu_{\tau}\rightarrow \nu_{e}+\gamma$, at the
temperature, they are at the same order, but while the temperature
further increases the radiative decay has a larger rate. The concrete
temperature also depends on the density, but the trend is obvious.
As Babu et al. pointed out that the window for MeV neutrino might be closed
\cite{Babu},
if so as aforementioned, $m_{\nu_{\tau}}$ gets to keV order, its lifetime
would become as troublesome as for $\nu_{\mu}$ of 160 keV.

(iii) The CP asymmetry.

At zero temperature, it is known that the penguin diagram  induces
a CP violation which can be observed at B- and D-decays. As estimated,
the absorptive angle which is equivalent to the strong scattering phase,
is relatively large. It is natural to ask if it can result in a CP
non-invariance at finite temperature and density which exist in the
early universe.

Since $\nu_{\tau}\rightarrow\nu_e+\gamma$ does not possess an absorptive
phase because $q^2=0$, it cannot induce a CP violation. Whereas, for
$\nu_{\tau}\rightarrow e^+e^-\nu_e$, there indeed exists an absorptive
angle, but there is the tree amplitude which is overwhelming at zero
temperature. Therefore, even though the absorptive angle is large, the CP
violation is very small because the tree contribution exists at the
denominator in expression (\ref{asym}).
When the temperature is very high, the penguin contributions
are much enhanced and can reach the same order as the tree contribution,
however, it is observed that
only the dispersive part increases, while the absorptive part remains
at the same order. Thus the absorptive angle $\theta_e$ turns out to be
rather small and the total CP violation at high temperature and density
keeps the same order as at zero temperature and remains to be small.
Therefore, even though the decay can serve as one of the CP non-invariance
sources, but definitely is not the main one to explain the matter dominance
over anti-matter.\\

\noindent {\bf Acknowledgment}
We would like to thank prof. R.K. Su for helpful discussions.

\vspace{2cm}

\centerline{\bf Figure Captions}

\noindent Fig.1 The Feynman diagrams for $\nu_{\tau}\rightarrow e^++e^-+
\nu_e$, where (a) is the tree diagram.

\noindent Fig.2 The dependence of the decay rate on the chemical potential,
the three curves correspond to $T=1\times 10^{10}$K (solid), 
$T=5\times 10^{11}$K (dashed), and $T=4\times 10^{12}$K (dotted), 
respectively.

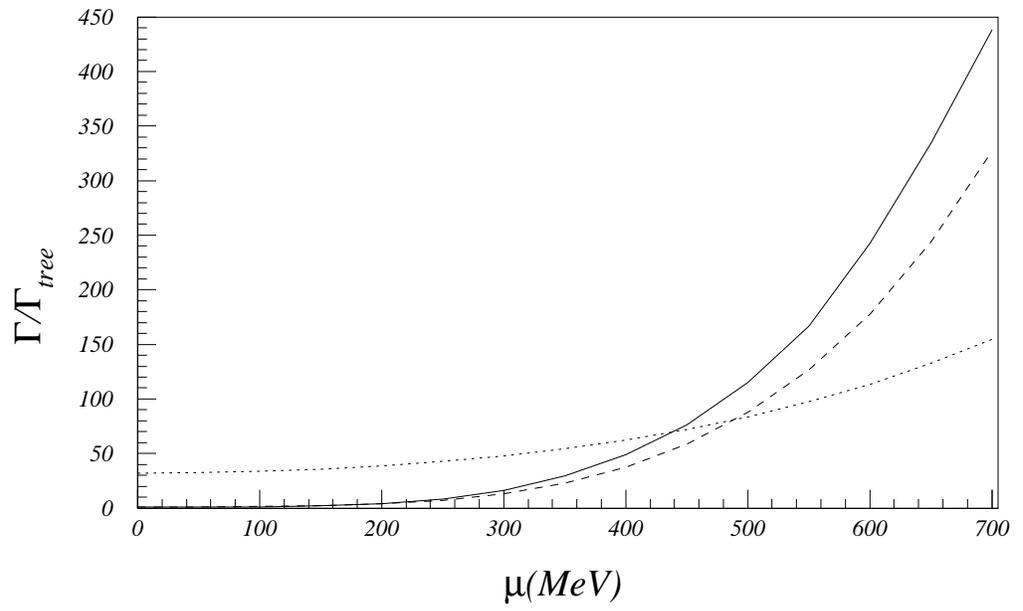
\end{document}